\documentstyle[preprint,aps,epsf]{revtex}

\newcommand{\beq}{\begin{equation}}
\newcommand{\eeq}{\end{equation}}

\newcommand{\SS}{Shastry-Sutherland}

\newcommand{\jj}{$J_{1}\!-\!J_{2}$}

\newcommand{\joj}{J_{2}/J_{1}}

\newcommand{\vk}{\vec{k}}
\newcommand{\vK}{\frac{\vQ-\vQo}{2}}
\renewcommand{\vr}{\vec{r}}
\newcommand{\vS}{\vec{S}}
\newcommand{\vQ}{\vec{Q}}
\newcommand{\vQo}{\vec{Q_0}}

\newcommand{\biup}{b_{i \uparrow}^{\dagger}}
\newcommand{\bidp}{b_{i \downarrow}^{\dagger}}
\newcommand{\bjup}{b_{j \uparrow}^{\dagger}}
\newcommand{\bjdp}{b_{j \downarrow}^{\dagger}}
\newcommand{\biu}{b_{i \uparrow}}
\newcommand{\bid}{b_{i \downarrow}}
\newcommand{\bju}{b_{j \uparrow}}
\newcommand{\bjd}{b_{j \downarrow}}

\newcommand{\Aij}{A_{i j}}
\newcommand{\Aijp}{A_{i j}^{\dagger}}
\newcommand{\Bij}{B_{i j}}
\newcommand{\Bijp}{B_{i j}^{\dagger}}

\begin{document}

\bibliographystyle{plain}

\title{First order transition between magnetic order \\ and valence-bond order
in a 2D frustrated Heisenberg model}
\vskip0.5truecm
\author {{\sc Marc Albrecht} and {\sc Fr\'ed\'eric Mila}}
\vskip0.5truecm
\address{Laboratoire de Physique Quantique\\
Universit\'e Paul Sabatier\\
31062 TOULOUSE}
\maketitle

\begin{abstract}
We study the competition between magnetic order and valence bond order in a two
dimensional (2D) frustrated Heisenberg model introduced some time ago by
Shastry
and Sutherland ({\sc B. Sriram Shastry} and {\sc Bill Sutherland}, {\em
Physica}
108{\bf B},1069 (1981) ) for which a configuration of dimers is known to be the
ground state in a certain range of parameters.
Using exact diagonalisation of small clusters, linear spin wave
theory and Schwinger boson mean field theory, we show that the
transition between the two types of order is first-order,
and that it takes place inside the domain where magnetic long-range
order is stable with respect to quantum fluctuations.
\end{abstract}


\newpage


Frustrating two-dimensional quantum magnets can yield new and exciting physics.
The possible effects include: i) an increase of quantum fluctuations that can
lead to the disappearance of long-range magnetic order; ii) the appearance of
new types of ground-states. While these effects are somehow compatible - they
both lead to the destruction of the usual type of order - the interplay
between them is not well understood. Let us be more specific and consider for a
moment a model that has been extensively studied in the past few years, namely
the \jj\ model~\cite{schulz}. Linear spin-wave theory (LSWT) predicts that
magnetic
order is destroyed around $J_2/J_1=1/2$ due to large quantum
fluctuations~\cite{doucot}. Consistent with this prediction,
series expansions for $S=1/2$ suggest that  a kind of
valence-bond order appears in that parameter range~\cite{gelfand}. Finally,
numerical
simulations~\cite{poilblanc} on this model suggest that the magnetization
vanishes continuously with increasing frustration, leading to the conclusion
that the phase transition between N\'eel order and valence-bond order is second
order. While the overall picture is
appealling, this conclusion is very surprising
because the degeneracy of the ground state is not the same in these phases (no
degeneracy for N\'eel order, four-fold degeneracy for valence bond order), and
one would have expected a transition due to a level crossing, in which case the
smooth disappearance of N\'eel order at the transition is a coincidence.
Besides, according to the Schwinger boson mean-field theory (SBMFT), which is
in principle an improvement over LSWT, quantum fluctuations are never strong
enough
to destroy long-range order~\cite{mila91}. So, whether magnetic long-range
order
has to disappear before another type of order can be stabilized is still an
open
question. As far as the \jj\ model is concerned, it seems difficult at the
moment
to go further because each technique is limited to a certain type of order. For
instance, the presence of valence-bond order has not been identified by
numerical
diagonalization of small clusters.

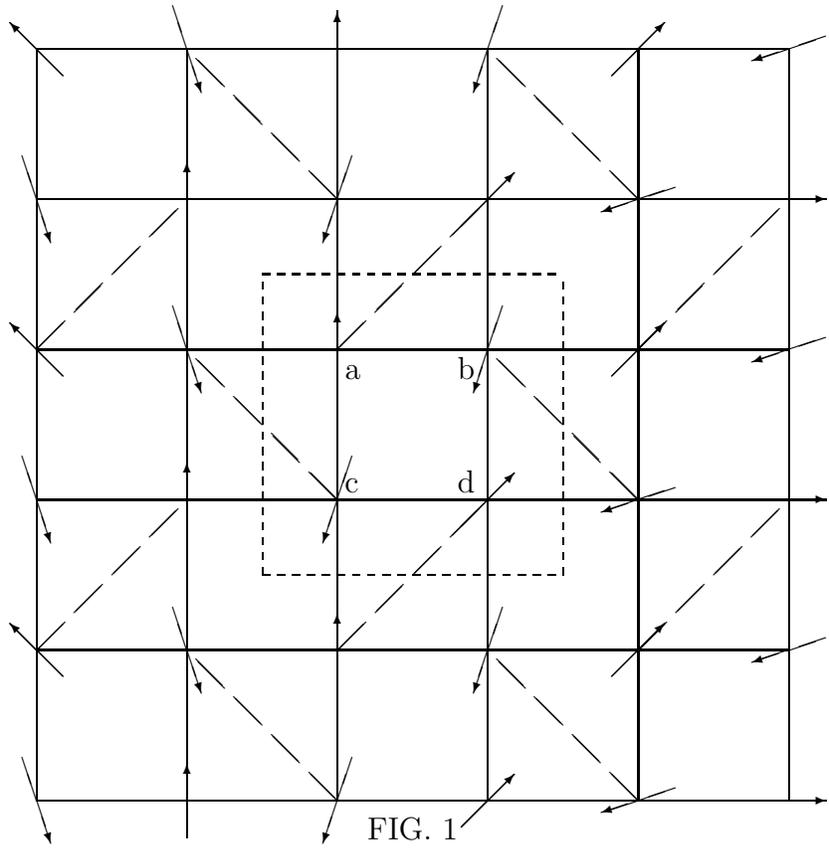
\begin{figure}
\caption{Shastry-Sutherland model. Full lines: links of magnitude $J_{1}$;
Long dashed lines: links of magnitude $J_{2}$. Short dashed lines: unit cell.
Arrows: typical helical configuration.}
\label{fi:reseau}
\end{figure}

In this paper, we propose to address that question by turning to another
model introduced some time ago by Shastry and Sutherland~\cite{ss}.
This model is a 2D frustrated Heisenberg model
defined by the Hamiltonian:
\beq
 H= J_{1} \sum_{<ij>}\vS_{i}.\vS_{j} + J_{2} \sum_{<lm>} \vS_{l}.\vS_{m}
\eeq
where $<\!\!ij\!\!>$ means pairs of nearest neighbors and $<\!\!lm\!\!>$
means pairs of
next-nearest neighbors linked by a dashed line in Figure~\ref{fi:reseau}. The
main
interest of this model is that
a valence bond solid has been shown to be the ground state for certain values
of
the parameters.  For such a model, the problem reduces to one question: do
quantum fluctuations destroy LRO before (or possibly when) the parameters
are such that the products of singlets becomes the ground state or do we
still have LRO at the transition?


Let us start with a review of the exact results obtained by Shastry and
Sutherland.
In the classical limit, when $S \rightarrow \infty$, the ground state is N\'eel
ordered if $ J_{1} < J_{2} $ and is helical otherwise. One can easily
show, by minimizing the energy of a single triangle, that the twist between one
spin and its nearest neighbor is given by $\theta = \pi \pm \arccos \left(
J_{1}/J_{2}  \right)$ and that the three spins must be coplanar. Once two
neighbour triangles are in this ground state, one can construct in a unique
way  such a pattern  on all the  other triangles of the lattice, thus building
a helix. One can show that this helix can be directed in four different
directions ( $ (0,\pm 1)$ and $(\pm 1,0)$ ) depending on the choice of the
values of $\theta$ for the  first two triangles. This adds a discrete four-fold
degeneracy to the ground state which  is also obviously continuously degenerate
because of the isotropy of the Hamiltonian.   A helix pointing towards $O_{x}$
is
represented on Figure~\ref{fi:reseau}.

In the quantum case, the state defined by $|\varphi\!\! > = \prod_{lm}
[l,m]$, where $[l,m]$ is the singlet state of the diagonal pair $(l,m)$, is an
eigenstate of  $H$ with an energy per site $E_{0} = -S(S+1)J_{2}/2J_{1}$. Using
a
variational principle, one can prove that this state is the ground state if $
\joj > 2 $ for  $ S=\frac{1}{2} $ and $ \joj  > 2(1+S) $ for $  S \geq 1 $.
This state has
been called a quantum spin liquid by Shastry and Sutherland because of the lack
of long range order. This exact result is the main interest of this
model. All these properties are summarized on the phase diagram of
Figure~\ref{fi:diag}.


To study quantum fluctuations, it is natural to start with linear spin wave
theory (LSWT)~\cite{holstein}.
The Hamiltonian being quadratic at this level of approximation,
one can easily determine the spectrum, and, from it, calculate the
magnetization from $ <\!\!S^{z}\!\!>=S- <\!\!a^{\dagger}_{i}a_{i}\!\!>$ in the
plane
$(\joj,1/S)$. The curve of stability
of the N\'eel ordered phase has the same shape as for the \jj\ and the
$J_{1}\!-\!J_{3}$ models: It drops at the critical value of the frustration
for which N\'eel order disappears~\cite{doucot,locher}.
However, the results for the helical phase are quite surprising: LSWT predicts
that this phase is unstable under quantum fluctuation for arbitrary large
values of $S$. This is due to a line of modes of zero frequency
defined by~: $ \cos^{2} k_{x} + \cos^{2} k_{y} = 1+ \frac{J_{1}}{J_{2}} $.
This makes the corrections to the staggered magnetization divergent in 2D.
However, this result must be viewed as an artifact of the method and not as
a meaningful  result: We know that the ground state is four-fold
degenerate, so the exact spectrum can not vanish on a continuous line of
wave-vectors.
This paradox can be lifted as follows: The dispersion given by LSWT corresponds
to modes of vanishingly small amplitudes in the classical limit. But for a
quantum system, the amplitudes of the local fluctuations cannot be arbitrarily
small, and one should recover a finite frequency for these modes.
A similar effect has
actually been observed in the collinear phase of the \jj\
model by Chandra {\em et al.}~\cite{chandra}.


As shown by these authors, a good method to take this effect
into account is to use the
Schwinger boson mean field theory (SBMFT)~\cite{arovas}.
This method starts from a representation of the spin algebra in terms of
bosonic operators: $ \vS_{i} = \frac{1}{2}
b_{i\sigma}^{\dagger}  \vec {\sigma}_{\sigma \sigma'} b_{i\sigma'} $, the size
of the spin being fixed by a constraint on the number of particles on each
site:
$b_{i\uparrow}^{\dagger} b_{i\uparrow}  + b_{i\downarrow}^{\dagger}
b_{i\downarrow} =2S$.
Defining operators that are quadratic in terms of the bosonic operators by
 $2\Bijp = \biup \bju + \bidp \bjd $ and $2\Aijp = \biup \bjdp
- \bidp \bjup $, the Hamiltonian can be written
\beq
H = \sum_{(i,j)} J_{ij}(:\Bijp \Bij: - \Aijp \Aij)
\eeq
At the mean-field level, one introduces the following order paramaters:
$ <\!\Aijp\!> = 2 \alpha_{ij} $ and $ <\!\Bijp\!>  =  2 \beta_{ij} $
and the Hamiltonien is replaced with:
\beq
H_{MF}= \sum_{(i,j)} J_{ij}\left( \beta_{ij}  (\Bij+\Bijp)
                          - \alpha_{ij} (\Aij+\Aijp)
                            -\beta_{ij}^2 +\alpha_{ij}^2 \right)
\eeq
Finally, the local constraint is replaced by a global one and is
enforced only on the average through the addition to the Hamiltonian of a term
$ \mu \sum_{i} \left( \biup \biu + \bidp \bid -2S \right) $,
where the chemical potential $\mu$ plays the role of a
Lagrange parameter.
In order to describe long range helical order in this formalism, one has to
multiply each Bose operator by a phase factor~\cite{cgt}:
$ b_{i} \rightarrow b_{i}\exp(i\vQ.\vr_{i}/2) $, where $\vQ$ is the pitch of
the helix. This is equivalent to perform a local rotation of angle
$\vQ.\vr_{i}$
on each site around a uniform quantization axis $z$.
The order parameters are then given by:

\beq
\begin{array}{ccccc}
\alpha_{ij} & = & \exp(-i\vQ.\vr_{ij}/2) <\! \biup \bju \! > & +
               &  \exp(i\vQ.\vr_{ij}/2) <\! \bidp \bjd \! > \\
\beta_{ij} & = & \exp(i\vQ.\vr_{ij}/2) <\! \biu \bjd \! > &  -
               &   \exp(-i\vQ.\vr_{ij}/2) <\! \bid \bju \! >
\end{array}
\label{eq:order.param}
\eeq

while the constraints reads:
\beq
2S=<\! \biup \biu \! >+ <\! \bidp \bid \! >
\label{eq:spin}
\eeq

The expectation values $ <\! b^{\dagger}_{i \sigma} b_{j \sigma} \! >  $
and  $ <\! b_{i \sigma} b_{j \sigma'} \! >  $
can be calculated from $ H_{MF}$ in terms of $\alpha_{ij}$ and $\beta_{ij}$,
so equations~(\ref{eq:order.param},\ref{eq:spin}) are
a system of non-linear equations from
which $\alpha_{ij}$ and $\beta_{ij}$ can be calculated.
A LRO state is described by a Bose condensate of $S^{*}$
particles~\cite{mila91} which gives the long range part of the correlation
functions.
Solving equations~(\ref{eq:order.param},\ref{eq:spin}) for  an arbitrary value
of $\vQ$ and a given value of $S$ leads to a excitation spectrum
that vanishes at  three points of the Brillouin zone located at
$\vk=\vK$ and
$\vk=\vK\pm\vQo$, where $\vQo$ depends on $S$ and $\joj$, but not on $\vQ$.\
So the physical solution, {\em i.e.} the solution that has a
Goldstone mode at $ \vk=\vec{0}$, is obtained by choosing $\vQ=\vQo$.
The spin correlation functions
$<\!\vS_{i}.\vS_{j}\!> = |\alpha_{ij}|^2 - |\beta_{ij}|^2$ have then the
following long range behaviour:

\beq
<\!\vS_{i}.\vS_{j}\!> \sim  S^{*2} \cos  (\vQo.\vr_{ij})
\eeq

Note that $\vQo$ is not equal to the classical value. This theory thus
allows one to calculate in an easy way the renormalization of the pitch
of the helix due to quantum fluctuations.
Note also that the rotation we perform is quite different from the one
done in~\cite{chandra}  which led
to $\left( \sin \theta_{ij} \right) \vS_{i}\wedge\vS_{j}$ terms in the
Hamiltonian. They are difficult to treat because they cannot be
written in terms of $\Aij$ and $\Bij$ and  neglecting them
gives a spectrum that does not have the appropriate Goldstone modes.

Looking for a solution of the mean-field equations such that both
the Bose condensate and the spectrum vanishes, we find the critical
value of the spin. The phase diagram is depicted in  figure~\ref{fi:diag}.
Within SBMFT, the N\'eel phase is found to be stable against quantum
fluctuations in a larger part of the phase diagram than within LSWT. For
every physical value of the spin, we find a domain of frustration in which
the system sustains helical LRO. This is the main advantage of the SBMFT
over LSWT which is unable to describe helical LRO in the particular case
of the \SS\ model.
The transition between N\'eel and helical order is always second order and
takes place at a frustration slowly increasing from $\joj=1$ in
the classical case to  $\joj \approx 1.1$  for $S=1/2$.

For a given value of the spin,
the \SS\ model has generally two solutions at the mean-field level: In
addition to the LRO one which exists for small enough frustration,
there is a spin liquid solution
defined by: $ |\phi\!\!>= \prod_{<\!\!<i,j>\!\!>} \Aijp |0\!\!>$.
This solution
has a gap in the excitation spectrum and the correlation functions
vanish at long distance.
It is the equivalent, within SBMFT, of the dimerized state, which
is always an eigenstate, and it has exactly the same energy. That the
agreement between the energies is perfect is to a
certain extent fortuitous because the wave-functions are not the same:
$|\phi\!>$ does not represent a spin
configuration because it mixes  wave-functions with different
numbers of particles, which is allowed by our approximation because the
constraint is satisfed only on the average.

The physical picture that emerges from these results is that
for all the physical values of $S$, a first order transition takes place
between long range helical order
and the valence bond ground state, a transition at
which the magnetization drops abruptely to zero, and a finite gap in the
excitation spectrum opens. For spin $1/2$, this transition occurs
at $\joj \approx 1.65$ . An other indication that the
transition we observed is first order comes from
the coexistence of the two solutions of the mean
field equations on the both sides of the transition.

For spin one-half, we have performed exact diagonalisation on small
clusters using Lanczos
algorithm. Our results are summarized on figure~\ref{fi:ex.diag}. We have
performed finite size scaling from the 8,16 and 20 sites
clusters for the staggered
magnetization and the ground state energy following the formula given
in~\cite{poilblanc}. These results agree very well SBMFT for
$\joj < 1.1$ and $ \joj > 1.65 $.
This confirms the reliability of the approximation we used and
our conclusions on the behaviour of the \SS\ model.
We want to point out that exact diagonalisation on small cluster
gives no valuable results in the strongly frustrated regime $ 1.1 < \joj
< 1.65 $ because small clusters cannot sustain helical LRO. As a
consequence, it is impossible to  extract
information about the transition between LRO and valence bond order
from this technique.


In conclusion,
we predict that for any physical value of the spin, the model
undergoes a first order phase
transition between magnetic order and valence-bond order.
This is  quite different from the scenario  outlined in the introduction
for the \jj\ model and according to which the system loses its LRO continously
at the transition to a valence-bond type of order. More qualitatively,
our results for spin one-half
can be summarized as follows:
For $ \joj < 1.1 $, the system exhibits N\'eel order.
This type of order is replaced by helical order
around $\joj \approx 1.1 $.
Finally, for $ \joj  \approx 1.65 $, there is a  first order phase
transition between the helical ordered phase and the spin liquid.

Another lesson to be learnt from this work concerns the limitation of exact
diagonalisation to localize a transition between an ordered state and a spin
liquid. From our study of the \SS\ model for $S=1/2$, one would have been
tempted, on the basis of exact diagonalisation alone,
to conclude that the transition is second order and takes place
at $\joj \approx 1.5$.
However, the SBMFT provides convincing arguments that this is an artifact
of this method because helical order cannot be realized on
small cluster.
In the case of the regular \jj\ model mentioned in the introduction,
helical order is also a good candidate around $\joj = 1/2$. At the classical
level a whole class of helical states are actually degenerate. Whether the
second order transition inferred from exact diagonalisation  is a real one
can then be questionned on the basis of our results. Work along these lines
is in progress.

We acknowledge useful discussions with T. Ziman and D. Poilblanc.
One of the
authors (F. M.) is particularly grateful to Christoph Bruder from whom he
learnt about this model.



\begin{figure}
\caption{Phase diagram of the Shastry-Sutherland model.}
\label{fi:diag}
\end{figure}

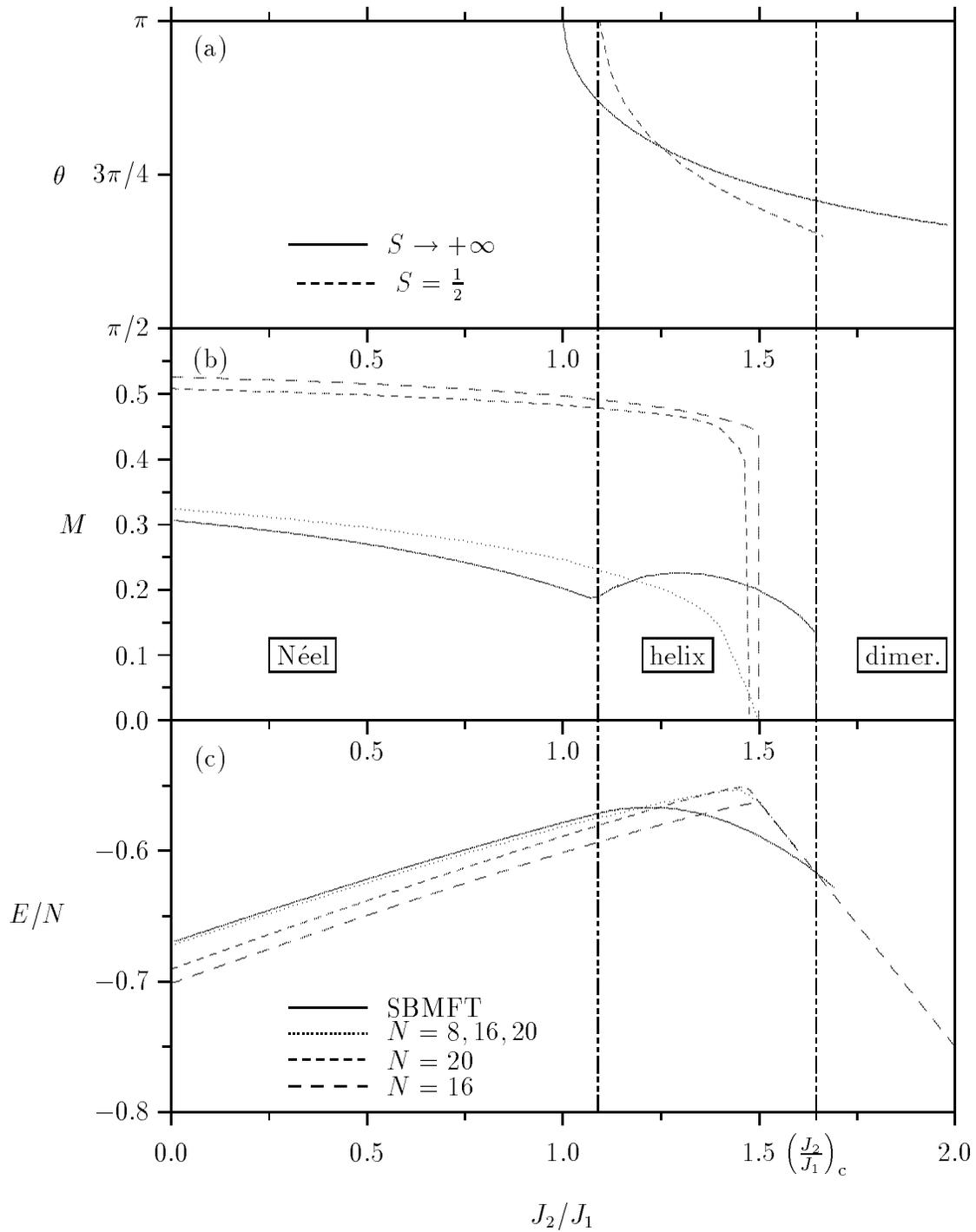
\begin{figure}
\caption{Variation of several quantities as a function of $\joj$ for $S=1/2$.
(a) Pitch of the helix; (b) staggered magnetisation; (c) ground state
energy.}
\label{fi:ex.diag}
\end{figure}


\newpage

\setlength{\unitlength}{2cm}
\begin{center}
\begin{picture}(5,5)(0,0)

\multiput(0,0)(0,1){6}{\line(1,0){5}}
\multiput(0,0)(1,0){6}{\line(0,1){5}}
\multiput(0,1)(2,0){3}{\begin{picture}(1,1)(0,0)
                    \multiput(0,0)(0.25,0.25){4}{\line(1,1){.19}}
                    \end{picture}}
\multiput(0,3)(2,0){3}{\begin{picture}(1,1)(0,0)
                    \multiput(0,0)(0.25,0.25){4}{\line(1,1){.19}}
                    \end{picture}}
\multiput(2,0)(0,2){3}{\begin{picture}(1,1)(0,1)
                    \multiput(0,1)(-0.25,0.25){4}{\line(-1,1){.19}}
                    \end{picture}}
\multiput(4,0)(0,2){3}{\begin{picture}(1,1)(0,1)
                    \multiput(0,1)(-0.25,0.25){4}{\line(-1,1){.19}}
                    \end{picture}}

\multiput(1.5,1.5)(2,0){2}{\begin{picture}(1,1)(0,0)
                    \multiput(0,0)(0,.1){20}{\line(0,1){.05}}
                    \end{picture}}
\multiput(1.5,1.5)(0,2){2}{\begin{picture}(1,1)(0,0)
                    \multiput(0,0)(.1,0){20}{\line(1,0){.05}}
                    \end{picture}}

\put(2.05,2.8){a}
\put(2.05,2.05){c}
\put(2.8,2.05){d}
\put(2.8,2.8){b}

\multiput(2,1)(0,2){3}{\vector(0,1){.25}}
\multiput(1,0)(0,2){3}{\vector(0,1){.25}}
\multiput(2,0)(0,2){3}{\vector(-1,-3){.096}}
\multiput(3,1)(0,2){3}{\vector(-1,-3){.096}}
\multiput(0,0)(0,2){3}{\vector(1,-3){.096}}
\multiput(1,1)(0,2){3}{\vector(1,-3){.096}}
\multiput(3,0)(0,2){3}{\vector(1,1){.177}}
\multiput(4,1)(0,2){3}{\vector(1,1){.177}}
\multiput(4,0)(0,2){3}{\vector(-3,-1){.246}}
\multiput(5,1)(0,2){3}{\vector(-3,-1){.246}}
\multiput(5,0)(0,2){3}{\vector(1,0){.25}}
\multiput(0,1)(0,2){3}{\vector(-1,1){.177}}

\multiput(1,0)(0,2){3}{\line(0,-1){.25}}
\multiput(2,0)(0,2){3}{\line(1,3){.096}}
\multiput(3,1)(0,2){3}{\line(1,3){.096}}
\multiput(0,0)(0,2){3}{\line(-1,3){.096}}
\multiput(1,1)(0,2){3}{\line(-1,3){.096}}
\multiput(3,0)(0,2){3}{\line(-1,-1){.177}}
\multiput(4,1)(0,2){3}{\line(-1,-1){.177}}
\multiput(4,0)(0,2){3}{\line(3,1){.246}}
\multiput(5,1)(0,2){3}{\line(3,1){.246}}
\multiput(0,1)(0,2){3}{\line(1,-1){.177}}
\end{picture} \\
FIG. 1 \\
\end{center}




\begin{references}


\bibitem{schulz} See e.g.  {\sc H. J. Schulz, T. A. L. Ziman} and
                 {\sc D. Poilblanc}, ``Frustration in
              Quantum Antiferromagnets'' in {\it Magnetic systems with
              competiting interactions} ed.
              H. Diep, World Scientific, Singapore (1994), and references
              therein.

\bibitem{doucot} {\sc P. Chandra} and {\sc B. Dou\c cot}, {\em Phys. Rev. B},
{\bf 38}, 9335 (1988).

\bibitem{gelfand} {\sc M.P. Gelfand, R.R.P. Singh} and {\sc D.A. Huse}
 	          {\em Phys. Rev. B} {\bf 40},10801 (1989)

\bibitem{poilblanc} {\sc H.J. Schulz} and  {\sc T.A.L. Ziman},
                  {\em Europhys. Lett.}, {\bf 18}, 355 (1992);
                {\sc H.J. Schulz, T.A.L. Ziman} and {\sc D. Poilblanc}
                {\em Phys. Rev. B} unpublished.


\bibitem{mila91} {\sc F. Mila, D. Poilblanc} and {\sc C. Bruder},
		 {\em Phys. Rev. B} {\bf 43},7891 (1991);
		 {\sc J. H. Xu} and {\sc C. S. Ting}, {\em Phys. Rev. B},{\bf
		 42}, 6861 (1990);
		 {\sc C. Bruder} and {\sc F. Mila},
	 	 {\em Europhys. Lett.}, {\bf 17},463 (1992).


\bibitem{ss} {\sc B. Sriram Shastry} and {\sc Bill Sutherland},
	    {\em Physica} 108{\bf B},1069 (1981)

\bibitem{holstein} {\sc T. Holstein} and {\sc H. Primakoff}
                  {\em Phys. Rev.} {\bf 58},1098 (1940)

\bibitem{locher} {\sc P. Locher} , {\em Phys. Rev. B} {Phys. Rev. B} {\bf
41}, 2537 (1990).

\bibitem{chandra} {\sc P. Chandra, P. Coleman} and {\sc A.I. Larkin},
                  {\em J. Phys. Condensed Matter} {\bf 2},7933 (1990)

\bibitem{arovas} {\sc D. P. Arovas} and {\sc A. Auerbach}
                 {\em Phys. Rev. B} {\bf 38},316 (1988)

\bibitem{cgt} {\sc H. A. Ceccatto, C. J. Gazza} and {\sc A. E. Trumper},
              {\em Phys. Rev. B} {\bf 47},12 329 (1992)


\end{references}
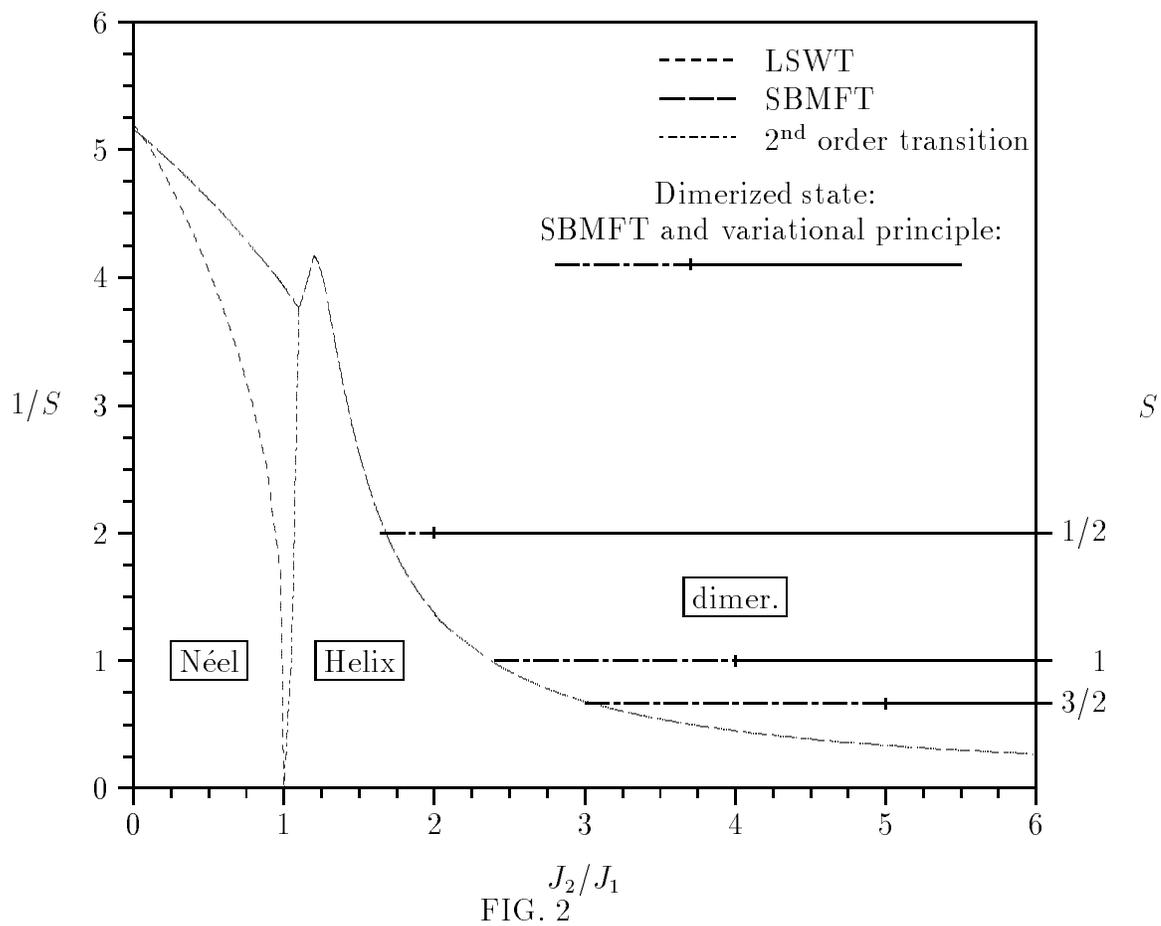
\end{document}